\documentclass[aps,pra,reprint,amsmath,amssymb,floatfix]{revtex4-1}

\usepackage{bm,graphicx,hyperref}
\hypersetup{
	colorlinks=true,
	linkcolor=cyan,      
	urlcolor=cyan,
	citecolor=cyan
}
\graphicspath{
{Figures/}
{/u/fs1/snms2/Documents/Work/Paper/Figures/}
}

\newcommand{\la}{\langle}
\newcommand{\ra}{\rangle}
\newcommand{\kk}{\mathbf{k}}
\newcommand{\GG}{\mathbf{G}}

\newcommand{\rr}{\mathbf{r}}
\newcommand{\zz}{\mathbf{z}}

\newcommand{\FF}{\mathbf{F}}

\newcommand{\RR}{\mathbf{R}}

\begin{document}
	
\title{Semiclassical Dynamics, Berry Curvature and Spiral Holonomy in Optical Quasicrystals}

\author{Stephen Spurrier}
\author{Nigel R. Cooper}

\affiliation{T.C.M. Group, Cavendish Laboratory, University of Cambridge,
	JJ Thomson Avenue, Cambridge, CB3 0HE, U.K.}

\date{\today}


\begin{abstract}
	We describe the theory of the dynamics of atoms in two-dimensional quasicrystalline optical lattices. We focus on a regime of shallow lattice depths under which the applied force can cause Landau-Zener tunneling past a dense hierarchy of gaps in the quasiperiodic energy spectrum. We derive conditions on the external force that allow for a ``semiadiabatic'' regime in which semiclassical equations of motion can apply, leading to Bloch oscillations  between the edges of a pseudo-Brillouin-zone. We verify this semiclassical theory by comparing to the results of an exact numerical solution. Interesting features appear in the semiclassical dynamics for the quasicrystal for a particle driven in a cyclic trajectory around the corner of the pseudo-Brillouin-zone: The particle fails to return to its initial state, providing a realization of a ``spiral holonomy'' in the dynamics. We show that there can appear anomalous velocity contibutions, associated with nonzero Berry curvature. We relate these to the Berry phase associated with the spiral holonomy, and show how the Berry curvature can be accessed from the semiclassical dynamics. Finally, by identifying the pseudo-Brillouin-zone as a higher genus surface, we show that the Chern number classification for periodic systems can be extended to a quasicrystal, thereby determining a topological index for the system.
\end{abstract}

\maketitle


\section{Introduction}

Quasicrystals \cite{shechtman84metallic,levine84quasicrystals} are an
interesting class of materials, in which the delicate mix of long-range order and lack of translational symmetry provides a setting that
is intermediate between periodic and random
systems \cite{steinhardt1987physics,sokoloff85unusual,poon92electronic}.
Recent work has shown that quasicrystals can lead to unconventional
dynamical \cite{sanchez-polencia05bose-einstein,lye07effect,roati08anderson,schreiber15observation,bordia2017probing}
and topological \cite{kraus12topologicalstates,bandres16topological,flicker15quasiperiodicity,kraus13fourdimensional,kraus2012topologicalequivalence,satija13chern}
properties.
Novel experimental settings have allowed these properties to
be explored with an unparalleled level of control in recent
years \cite{vardeny2013optics,sanchez-polencia05bose-einstein,steurer07photonic}
compared to conventional condensed matter systems. A particularly
flexible setting in which quasicrystals have begun to be
studied \cite{guidoni1997quasiperiodic,guidoni1999atomic,roati08anderson,schreiber15observation}
is in ultracold
gases \cite{bloch2005ultracold,bloch2012quantum,lewenstein07ultracold}. Here
the interference pattern from overlapping laser beams can generate a
wide variety of potential
landscapes \cite{petsas94crystallography,becker2010ultracold,tarruell2012creating,jo2012ultracold}
-- referred to as optical lattices -- including a variety of
one-dimensional (1D) \cite{roati08anderson,singh15fibonacci} and
two-dimensional
(2D) \cite{guidoni1997quasiperiodic,jagannathan13eightfold}
quasicrystals, that are essentially free from disorder and also highly
tunable.

The lack of disorder in optical lattices offers an advantage over
solid state in allowing for the study of phase coherent transport
phenomena without scattering \cite{chien2015quantum}. The classic
example is the demonstration of Bloch oscillations in an optical
lattice \cite{bendahan96bloch}, a phenomenon which has not been
observed for bulk crystalline electrons. The theory that describes
these phenomena is semiclassical dynamics \cite{chang95berry}. This
says that under the influence of a weak external force a particle's
motion is determined by the band structure and by the geometrical
properties of its eigenstates encoded in the Berry
curvature \cite{berry84quantal} -- a quantity that is intimately
related to the topological properties of the band
structure \cite{thouless82quantized,simon83holonomy,kohmoto1985topological}. The
ability to access these properties cleanly in cold
atoms \cite{price12mapping} has been exploited experimentally to
measure geometrical and topological features of energy bands of
fundamental models \cite{aidelsburger2015measuring,jotzu2014}.

Here we explore the nature of semiclassical dynamics in an optical
quasicrystal. We develop this for lattices of shallow depth,
corresponding to the nearly free electron limit of solid-state
terminology.  Our approach exploits the idea that within this limit,
and due to the quasiperiodicity, there is an unending fractal hierarchy
of gaps in the band structure controlled by perturbation
theory \cite{lu1987electronic}. For any non-zero external force,
Landau-Zener tunneling will make only a finite number of these
gaps relevant within the semiclassical
dynamics \cite{zhang15disruption}. The resulting theory is closely
analogous to that of a periodic system except that the unconventional
rotational symmetries -- disallowed for periodic systems -- can lead
to exotic band structures. As a surprising result of this, we find a
realization of a spiral holonomy \cite{cheon09new,cheon09exotic}, involving a
permutation between bands under an adiabatic cyclic trajectory. This
phenomena is a generalization of Berry's phase \cite{berry84quantal}
and the Wilczek-Zee holonomy \cite{wilczek84appearance}. A comparison
against an exact solution to the time-dependent Schr{\" o}dinger
equation verifies that the semiclassical theory works well within the
shallow-lattice limit.  We show under what conditions Berry curvature
effects can appear for semiclassical dynamics in quasicrystals, at
least within the shallow-lattice limit. Finally we discuss how these
ideas are generalized to arbitrary rotational symmetries.


\section{Model\label{sec:model}}
	
\begin{figure}[hbtp]
	\centering
	\includegraphics[width=.9\linewidth]{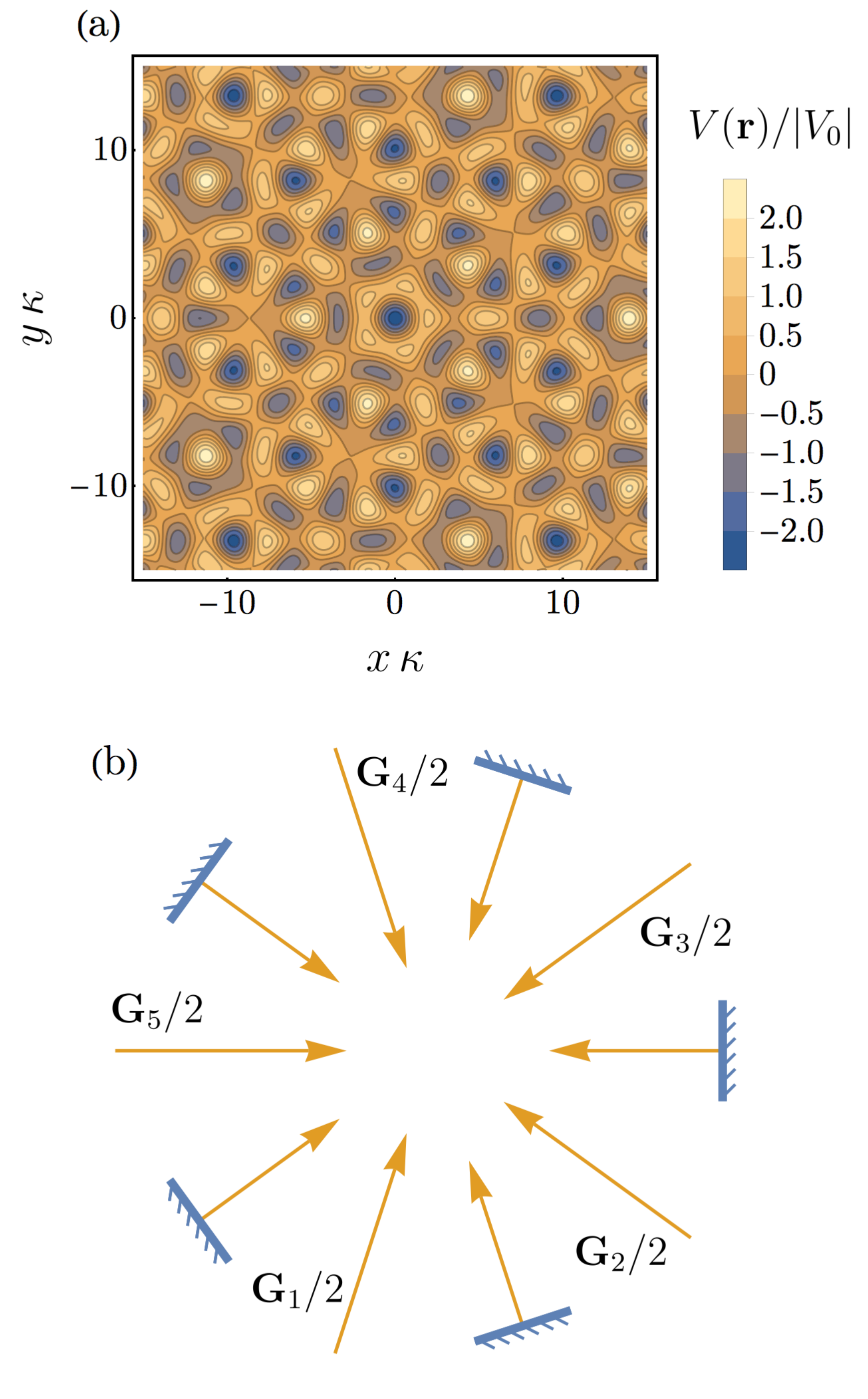}%
	\caption{\label{fig:potentiallaserconfig}(a) The considered quasiperiodic optical lattice potential given by Eq. (\ref{eq:V}) with  $ V_0 < 0 $ and $ \theta_i = \pi/10 $ for $ i = 1, \dots , 5$. (b) A fivefold arrangement of mutually incoherent beams with wave vectors $ \GG_i/2 $ plus coherent reflections. The imposed fivefold rotational symmetry forces the optical lattice potential to be quasiperiodic because a fivefold symmetry is disallowed in periodic systems.
}
\end{figure}

We consider a two-dimensional optical lattice quasicrystal shown in Fig. 
\ref{fig:potentiallaserconfig}(a), with
potential,
\begin{align}
V(\rr) \equiv  \frac{V_0}{2}\sum_{j=1}^{5} \cos ( \GG_{j} \cdot \rr +\theta_{j}), \label{eq:V}
\end{align}
where $ V_0 $ sets the overall strength of the potential, $ \GG_j $ are wave vectors given by
\begin{align}
\GG_j \equiv 2\kappa \left(\cos(2\pi j/5),\sin(2\pi j/5)\right),
\label{eq:GG}
\end{align}
and $ \theta_j $ are arbitrary phase offsets. This optical lattice could be generated using standard experimental methods using a laser arrangement shown in Fig.~\ref{fig:potentiallaserconfig}(b), consisting of five mutually incoherent laser standing waves set at an angle of $ 2 \pi /5 $ with respect to one another.

\begin{figure}[tb]
	\centering
	\includegraphics[trim={0.45cm .0cm -1.cm .0cm}, clip,width=0.9\linewidth]{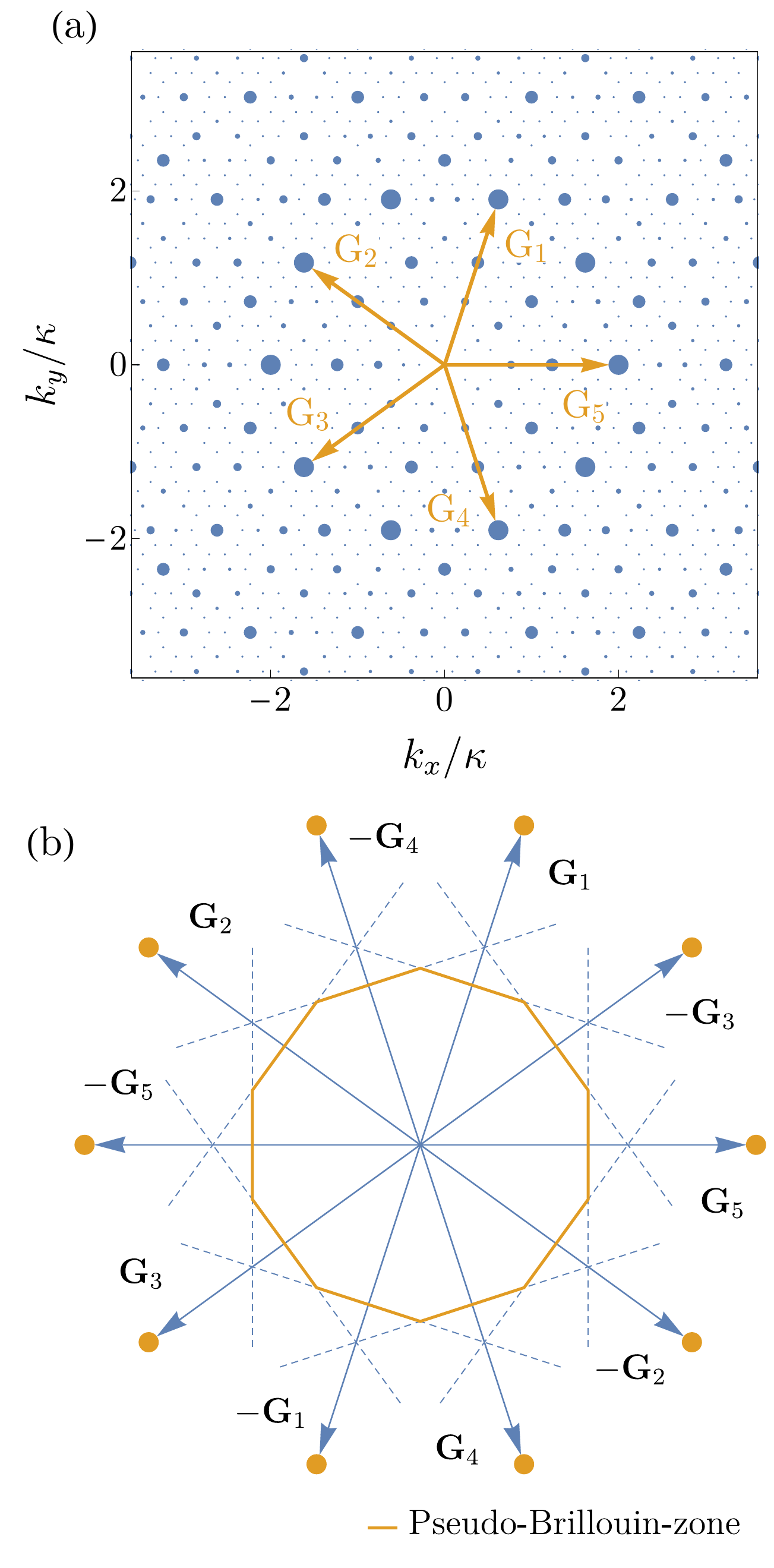}
	\caption{(a) The set of all combinations of the five principal wave vectors $ \GG_i $ -- referred to as the reciprocal lattice -- forms a dense set of points in $ k $-space. The corresponding set of plane wave states forms the basis for the eigenstates and in the shallow-lattice limit the free particle dispersion will develop a hierarchy of gaps proportional to the point sizes shown. (b) The largest gaps are those along the lines of degeneracy between the center and the 10 principal wave vectors $ \pm \GG_i $ which together form a decagonal boundary to a region referred to as the pseudo-Brillouin-zone.}
	\label{fig:reciplatPBZ}
\end{figure}

We highlight that this potential satisfies the definition of a quasicrystal \cite{steinhardt1987physics} in that the minimum number of basis vectors needed to span its Fourier transform (four) is more than the dimension of the space (two). These basis vectors can be chosen as any four of the five vectors $\GG_j$, Eqn.~(\ref{eq:GG}). (The reduction from five to four arises from the linear dependence $\sum_{j} \GG_j = 0$.)
In general, the eigenstates for the Hamiltonian, 
\begin{align}
\hat{H} = \frac{\hbar^2 \hat{\kk}^2}{2m} + V(\rr),
\label{eq:ham}
\end{align}
can be found by expanding in a basis of plane wave states $|\kk+\GG\rangle$ where
\begin{align}
\GG = \sum_{i} n_i \GG_i, \label{eq:fouriercomponents}
\end{align}
runs over all possible vectors formed from the four linearly independent basis vectors, as $n_i$ run over all integers. For crystalline lattices, $\GG$ forms the reciprocal lattice. For the quasicrystal, the key difference is that this set of vectors fills reciprocal space densely, as shown in Fig.~\ref{fig:reciplatPBZ}(a).

An important assumption we work with throughout the paper is the shallow-lattice limit,
\begin{align}
V_0 \ll E_R,
\label{eq:shallow}
\end{align}
where $ E_R \equiv \hbar^2 \kappa^2 / 2 m $ is the recoil energy. In this limit the band structure and eigenstates for the Hamiltonian (\ref{eq:ham})
can be found by applying perturbation theory.  Away from lines of degeneracy between free particle states (Bragg planes), the energy spectrum is given by
\begin{align}
E(\kk)  = \frac{\hbar^2 k^2}{2m}+ \sum_{\kk' \in \{\GG\}} \frac{|\la \kk | V | \kk' \ra |^2}{E_0(\kk)-E_0(\kk')},
\end{align}
and the effect of $ V $ is just a second-order correction. (We have used $\la \kk | V | \kk \ra=0$.) On the other hand, along any twofold degeneracy -- at the crossing of the free particle energies for $\kk$ and $\kk'$ say --  degenerate perturbation theory must be used. This opens a gap  proportional to the matrix element between the two degenerate states
\begin{align}
\Delta_\text{gap} = 2 |\la \kk | \hat{V} | \kk' \ra |,
\end{align}
with matrix elements given by the Fourier coefficients
\begin{align}
V_{\kk-\kk'} \equiv \la \kk | \hat{V} | \kk' \ra = \int d\rr \, V(\rr) e^{-i (\kk-\kk')\cdot\rr} .
\end{align}
The only nonzero Fourier coefficients, and therefore nonzero gaps to first order in $V$, are those shown in Fig.~\ref{fig:reciplatPBZ}(b) corresponding to $ \pm\GG_j $. These define a region known as the pseudo-Brillouin zone (PBZ) \cite{rogalev2015fermi,smith87pseudopotentials,lu88acoustic,burkov93optical}. 

These gaps represent Bragg scattering processes to first order in $V$. To higher orders of perturbation theory, gaps will open along all lines of degeneracy, corresponding to effective multiple scattering processes. Therefore the initial free particle dispersion develops a dense hierarchy of gaps \cite{lu1987electronic,rolof2013electronic,zhang15disruption}. However, in the shallow-lattice limit (\ref{eq:shallow}) these gaps in the hierarchy have rapidly decreasing sizes with order of perturbation theory. Thus, under suitable conditions, the hierarchy can be truncated in their contributions to physical observables. Indeed we make this idea explicit in the  ``semiadiabatic" limit which we now define, and which allows access to a description based on semiclassical dynamics. 

\section{Semiclassical Dynamics\label{sec:SCD}}

For ordinary periodic systems the equations of semiclassical dynamics play a fundamental role in our understanding of numerous transport properties \cite{ashcroftmerminsolidstateshysics,chang95berry,xiao10berry}. These allow for a reduction in information from an underlying quantum theory to a pair of classical equations requiring information about only the band structure and Berry curvature. In the setting of cold atoms, where there is little disorder and where scattering from interactions can be made weak, they can provide an accurate description of the dynamics over long times \cite{price12mapping,aidelsburger2015measuring}.

This theory describes the motion of a wave packet centered at $ \kk $ in reciprocal space and $ \rr $ in real space under the influence of an external force $ \FF $. In solid state systems this force arises from the electric or magnetic fields acting on the electron, whereas, because atoms are neutral, for ultracold atomic gases this force typically arises from tilting or accelerating the lattice. For a sufficiently weak external force, such that the typical evolution time is sufficiently long compared to the inverse of the gap, the wave function will remain in a single band throughout the evolution and the resulting dynamics will be accurately described by the semiclassical equations of motion \cite{chang96berry,xiao10berry},
\begin{align}
\dot{\kk}=&\frac{1}{\hbar}\mathbf{F},\label{eq:SC1}\\
\dot{\rr}=&\frac{1}{\hbar} \frac{\partial E ( \kk)}{\partial \kk}-(\dot{\kk}\times \hat{\mathbf{z}})\Omega(\kk).\label{eq:SC2}
\end{align}
The first equation describes the trajectory of $ \kk $ through reciprocal space under the external force $ \FF $. While the second relates the motion in real space to the dispersion relation $ E(\kk) $ \cite{ashcroftmerminsolidstateshysics} and an additional term \cite{karplus54hall} (often referred to as the anomalous velocity) proportional to the Berry curvature $ \Omega(\kk) $ defined by \cite{berry84quantal}
\begin{align}
\Omega(\kk)
\equiv \bm{\nabla}_\kk \times[i\langle u_\kk|\bm{\nabla}_\kk u_\kk \ra]\cdot\hat{\zz}, \label{eq:berrycurvature}
\end{align}
with $ u_\kk(\rr) \equiv e^{-i \kk \cdot \rr} \psi_\kk(\rr) $.

Applying these equations to a quasicrystal presents a number of difficulties. The central issue is the interpretation of $ \kk $. In a periodic system $ \kk $ is the crystal momentum and is thereby only defined up to the addition of a reciprocal lattice vector. This encourages one to restrict $ \kk $ to the Brillouin zone, ensuring that each $ \kk $ labels a unique eigenstate. A similar approach for quasicrystals is inappropriate as here the Brillouin zone is infinitesimally small (since there is no lower limit on the size of a reciprocal lattice vector). Instead throughout the following we essentially use a repeated zone scheme in which $ \kk $ is allowed to take any value in reciprocal space. 

\begin{figure}[tb]
	\centering
	\includegraphics[trim={0.1cm .0cm -0.1cm .0cm}, clip,width=1.\linewidth]{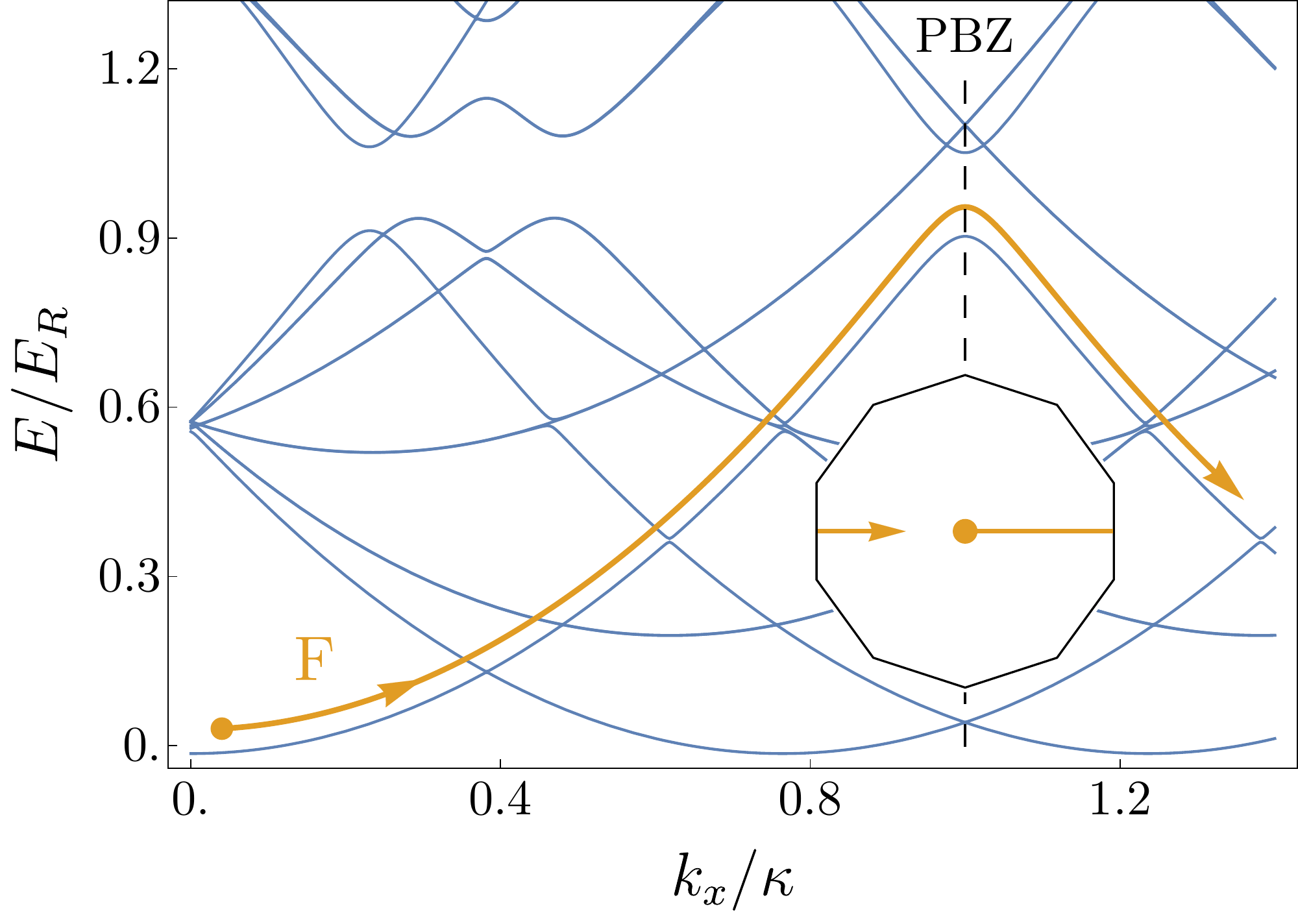}
	\caption{In the semiadiabatic limit (\ref{eq:semiadiabatic}), the only relevant gaps for semiclassical dynamics are those along the PBZ boundary. These are followed adiabatically, whereas the dense set of smaller gaps are eliminated by Landau-Zener tunneling. This idea is illustrated by the energy spectrum $ E(\kk) $ along a trajectory past the PBZ with a basis truncated to approximately 100 elements and $ V_0/E_R = 0.3 $.}
	\label{fig:LZ}
\end{figure}

Closely related to the issue of how to interpret $ \kk $ is the problem of defining $ E(\kk) $ and $ \Omega(\kk) $ for a quasicrystal. Our approach to this problem is twofold. First we exploit the shallow-lattice limit (as was presented in the preceding section), within which the spectrum simplifies to a free particle  dispersion which is broken into a dense hierarchy of gaps. Secondly we use the idea that under an external force all gaps with a size below a certain threshold will be essentially removed from the dynamics due to Landau-Zener tunneling. The Landau-Zener probability for tunneling through an avoided crossing between two free particle states, $ | \kk - \GG \ra $ and $ | \kk - \GG' \ra $, is given by \cite{landau1932theorie,*zener32non-adiabatic}
\begin{align}
	P_\text{LZ} = e^{-\alpha \Delta_\text{gap}^2/F},
\end{align}
with $ F = |\FF| $, $ \alpha = \pi m / 2 \hbar^2 \delta $ and $ \delta  = | \hat{\FF} \cdot (\GG-\GG') | $. For all gaps that satisfy
\begin{align}
	 \Delta_\text{gap}^2 \ll F/\alpha,
\end{align}
the probability of Landau-Zener tunneling will go to one, $ P_\text{LZ} \to 1 $, and these gaps will be essentially ignored in the semiclassical dynamics. If the force is also carefully chosen so that the dynamics remain adiabatic with respect to the remaining gaps, the dynamics will then be accurately described by the semiclassical equations of motion (\ref{eq:SC1}) and (\ref{eq:SC2}). With $ E(\kk) $ and $ \Omega(\kk) $ interpreted as the remaining part of the spectrum which is relevant in the semiclassical dynamics. 

It is important to stress that unlike periodic systems in which a band structure is always well defined, it is only within a dynamical picture, and within a certain window of external forces, that a particular effective band structure emerges. The connection between the dynamics and $ E(\kk) $ and $ \Omega(\kk) $ via semiclassical dynamics is therefore essential in defining these quantities for a quasicrystal. It should also be highlighted that semiclassical dynamics for a quasicrystal is more restrictive than for periodic systems. This is because we require both adiabaticity with respect to some gaps (as with periodic systems) and also non-adiabaticity for others (unlike periodic systems).

These ideas highlight that the particular semiclassical dynamics found in a quasicrystal will depend on the magnitude of the external force, with increasingly weaker regimes of force resulting in a growing number of gaps becoming relevant \cite{zhang15disruption}. Throughout the following we focus on a particularly simple regime of forcing which we refer to as the ``semiadiabatic limit.'' We define this as the regime in which the dynamics are adiabatic with respect to the largest gaps---those of order $ V_0 $ which form the boundary of the PBZ, but nonadiabatic with respect to the gaps of order $ V_0^2/E_R $ (as well as all smaller gaps in the hierarchy), as shown in Fig.~\ref{fig:LZ}. Therefore the dynamics are semiadiabatic when $ F $ satisfies
\begin{align}
\left(\frac{V_0}{E_R}\right)^4 \ll \frac{F}{\kappa E_R} \ll \left(\frac{V_0}{E_R}\right)^2 \label{eq:semiadiabatic}.
\end{align}

\begin{figure}[tb]
	\centering
	\includegraphics[trim={0.0cm 0.cm 0.cm 0.cm}, clip,width=0.99\linewidth]{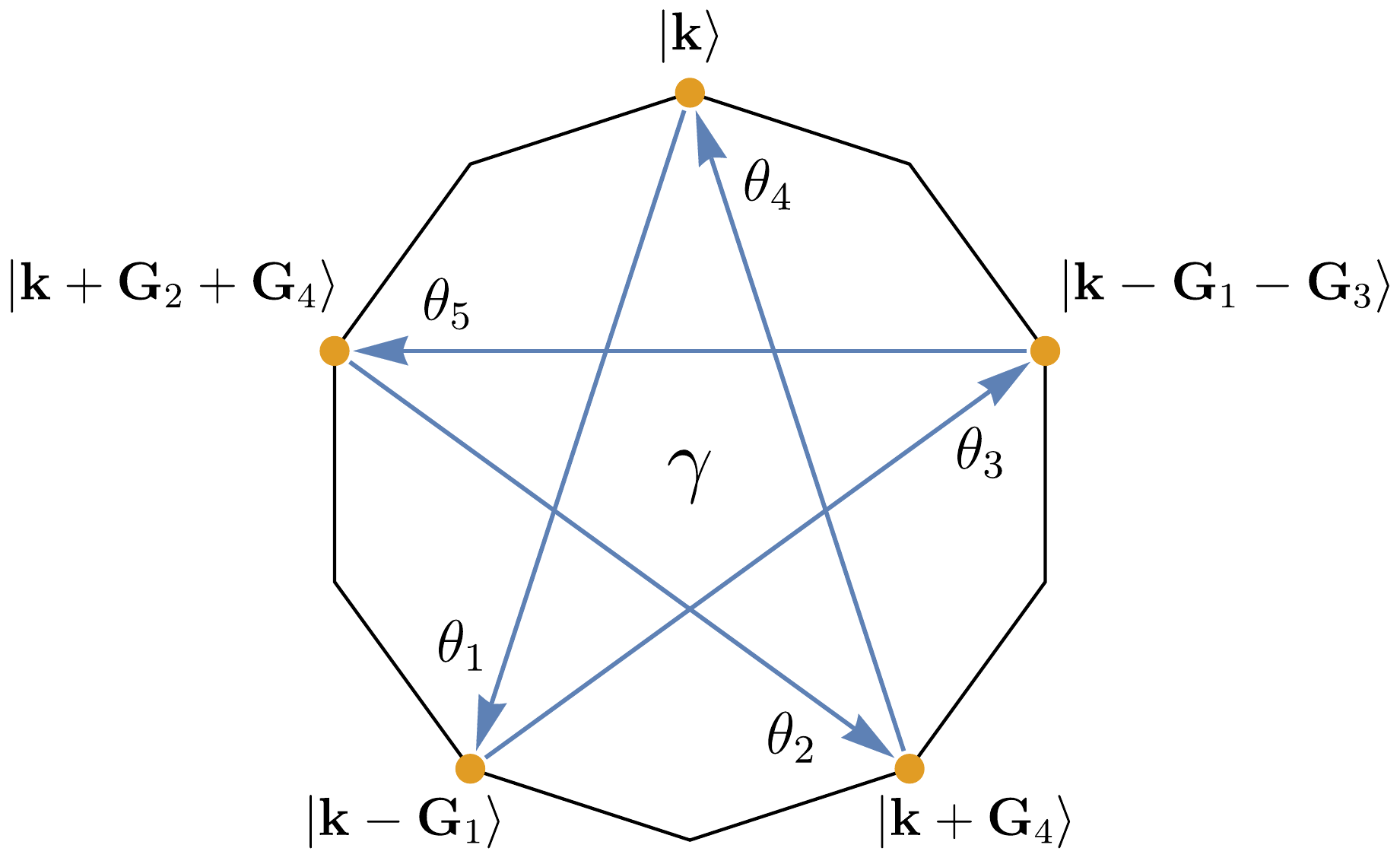}
	\caption{Local to the topmost corner the state $ |\kk\ra $ is coupled to four other states (each marked with a point), with the mixing between these described by the Hamiltonian $ H^\text{corner}_\kk $ as in Eq. (\ref{eq:Hcorner}). The off-diagonal couplings are represented by arrows and the corresponding phases $ \theta_i $ have been included. As discussed in Sec.~\ref{sec:berryphase} each phase is gauge dependent, however, since the couplings form a closed loop the total, $ \gamma $, is gauge invariant which allows for nontrivial Berry phase and curvature.
	}
	\label{fig:TotalPhase}
\end{figure}

The form of $ E(\kk) $ and $ \Omega(\kk) $ in the semiadiabatic limit falls into two cases depending on the location of $ \kk $ in the PBZ. Away from the boundary of the PBZ, $ V(\rr) $ has little effect and to leading order one has free particle dispersion $ E(\kk) = \epsilon_k $, with $ \Omega(\kk) $ zero. Whereas nearby the boundary, $ E(\kk) $ and $ \Omega(\kk) $ are determined by considering mixing between the free particle states that are degenerate there. Along a straight edge, this involves just two states, whereas at a corner we have the more interesting case of mixing between five degenerate states. These can be identified by considering a series of scatterings at a corner, as shown in Fig.~\ref{fig:TotalPhase}. For example, if we consider $ \kk $ nearby the topmost corner, the state $ |\kk \ra $ will be coupled to the states $ |\kk- \GG_1\rangle $ and $|\kk + \GG_4 \rangle $, and these to the states $ |\kk- \GG_1 - \GG_3 \rangle $ and $|\kk + \GG_2 + \GG_4 \rangle $ respectively, with the final two states coupled to each other. The Hamiltonian that describes the mixing between these five states is given by,
\begin{align}
&H^\text{corner}_\kk = \nonumber\\
&\begin{pmatrix}
\epsilon_\kk	&	V_{\GG_1}	& V_{-\GG_4}	& 0 & 0\\
V_{-\GG_1} &	\epsilon_{\kk-\mathbf{G_1}}	&	0	& V_{\GG_3}	& 0 \\
V_{\GG_4} &	0 &	\epsilon_{\kk+\mathbf{G_4}}	& 0	&	V_{-\GG_2}\\
0	& V_{-\GG_3}	&	0	&	\epsilon_{\kk-\mathbf{G_1}-\mathbf{G_3}}	&	V_{\GG_5}\\
0	&	0	& V_{\GG_2}	& V_{-\GG_5} & \epsilon_{\kk+\mathbf{G_2}+\mathbf{G_4}}
\end{pmatrix} \label{eq:Hcorner}
\end{align}
with $ V_{\GG_j}= (V_0/4) e^{i \theta_j} $. 

While we will focus on the semiadiabatic limit throughout the rest of the paper, essentially all the results we discuss can be simply extended to a regime in which the force is tuned to a different set of gaps. Generally if one chooses the force according to 
\begin{align}
\left(\frac{V_0}{E_R}\right)^{2(n+1)} \ll \frac{F}{\kappa E_R} \ll \left(\frac{V_0}{E_R}\right)^{2n},
\end{align}
the situation described for the semiadiabatic case is altered by replacing the set of principal wave vectors $ \GG_i $ with a set of 10 vectors $ \GG_i' $ associated with $ n $th order scatterings. This set is found by taking the smallest magnitude wave vectors from the set of all $ n $th order combinations of $ \GG_i $ (these will necessarily have the same tenfold symmetry), and will have phases $ \theta_i' $ equal to the sum of the $ n $ phases associated with the $ n $ wave vectors $ \GG_i $. One can then define a corresponding $ n $th order PBZ defined by the set of perpendicular bisectors to $ \GG_i' $, along with a similar matrix to $ H^\text{corner}_\kk $ in (\ref{eq:Hcorner}) describing the dynamics nearby a corner.



\section{Bloch Oscillations\label{sec:blochiscillations}}

An immediate result of the above discussion is that, within the semiadiabatic limit, a constant external force will drive Bloch oscillations in a manner closely analogous to those in periodic systems. The possibility of Bloch oscillations in a quasicrystal was first identified in a number of numerical studies \cite{diez96dynamical,sanchez-polencia05bose-einstein}. There the Bloch oscillations were found to be quasiperiodic whereas, within the semiadiabatic limit defined here, it is possible to have approximately periodic oscillations if the force is directed along certain high symmetry directions. For arbitrary directions, the resulting evolution can be highly complicated,  as indeed is also the case for periodic crystals \cite{price12mapping}. An interesting difference for quasicrystalline Bloch oscillations is that, as the force is reduced, new gaps in the hierarchy will become relevant and new Bloch oscillation periods will appear. This point will remain true for arbitrarily small forces and therefore quasicrystalline Bloch oscillations contain a much richer structure.

The prediction of Bloch oscillations can be used to test the validity of the semiadiabatic approximation by comparing against an exact numerical solution of the time-dependent Schr{\"o}dinger equation, which takes the form,
\begin{align}
i\hbar \,
\partial_t a_{\kk}=\epsilon_\kk a_{\kk}+\sum_{\GG_j} V_{\GG_j} a_{\kk-\GG_j},\label{eq:TDSE}
\end{align}
in a basis of free particle states,
\begin{align}
|\psi_{\kk}\rangle = \sum_{\GG \in \{\GG\}} a_{\kk-\GG} |\kk-\GG\rangle,
\end{align}
where the sum is over the reciprocal lattice, $ \epsilon_\kk \equiv \hbar^2 |\kk|^2/2m $ is the free particle dispersion, and $ V_{\GG_j} = (V_0/4) e^{i \theta_j} $ are the couplings due to the potential.

\begin{figure}[tb]
	\centering
	\includegraphics[trim={0.1cm .0cm 0.cm 0.cm}, clip,width=0.99\linewidth]{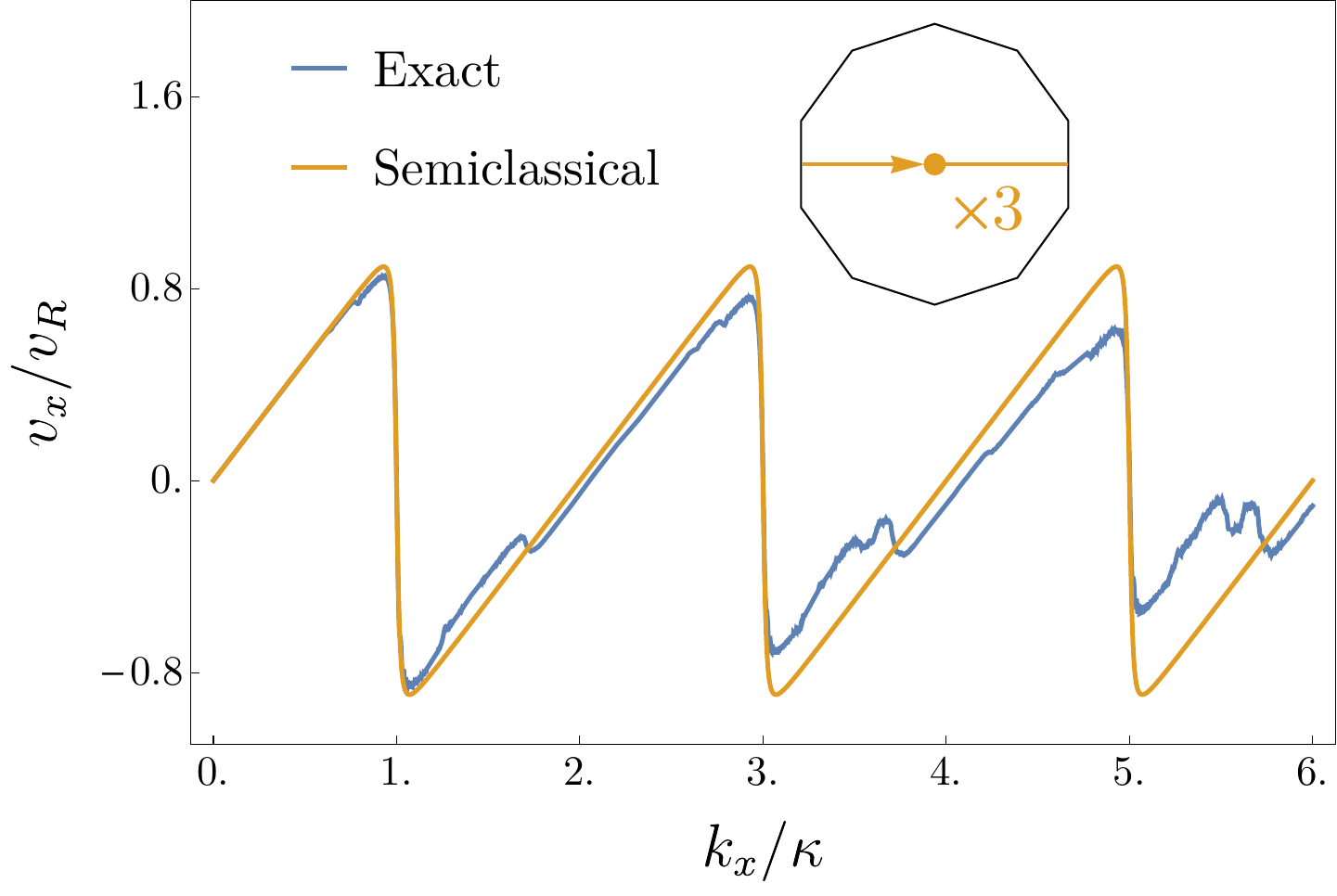}
	\caption{Comparison between the semiclassical approximation and the exact numerical result of the mean  velocity $ v_x $ for a trajectory along the high symmetry direction shown inset and with the parameter values $ V_0/E_R = 0.15 $, $ F/\kappa E_R = 3.6 \times 10^{-4} $, and where $ v_R \equiv \hbar \kappa /m $ is the recoil velocity. The results demonstrate approximately periodic Bloch oscillations, while the match between the exact and semiclassical results can be improved by going to smaller $ V_0/E_R $.}
	\label{fig:SCDkx}
\end{figure}

We have solved (\ref{eq:TDSE}) numerically, 
choosing our numerical basis large enough so that our results for the mean velocity have converged
for any given set of parameters. The comparison between these exact results and the semiclassical dynamics is shown in Fig.~\ref{fig:SCDkx}. 
The results of this comparison suggest that the semiclassical approximation should remain valid up to roughly $ V_0/E_R \approx 0.15 $, with the external force that satisfied the semiadiabatic limit the closest found to be $ F/\kappa E_R = 3.6 \times 10^{-4} $. Beyond this value of $ V_0/E_R $ the window of allowed values for $ F $ that satisfy (\ref{eq:semiadiabatic}) becomes so narrow that it becomes impossible to choose a single value that satisfies both limits adequately. The signal of this breakdown is the appearance of new oscillation frequencies corresponding to previously neglected higher order gaps.

Observing these Bloch oscillations experimentally requires that the relatively long evolution times needed within the semiadiabatic limit do not exceed the typical lifetimes of the atomic gases used, which are of the order of a few seconds. For the parameter values found numerically ($ V_0/E_R = 0.15 $ and $ F/\kappa E_R = 3.6 \times 10^{-4} $) the time $ T $ to complete a single Bloch oscillation is given by $ T  = 5 \times10^{3} \hbar/E_R $. For $^{23}$Na and ${}^{87}$Rb this takes the values of $ T \approx 0.02$s and $0.2 $s, and could therefore be quite challenging experimentally. 

\section{Spiral Holonomy\label{sec:spiralholonomy}}
\begin{figure*}[tbp]
	\centering
	\includegraphics[trim={.5cm .5cm .3cm .1cm}, clip,width=.85\linewidth]{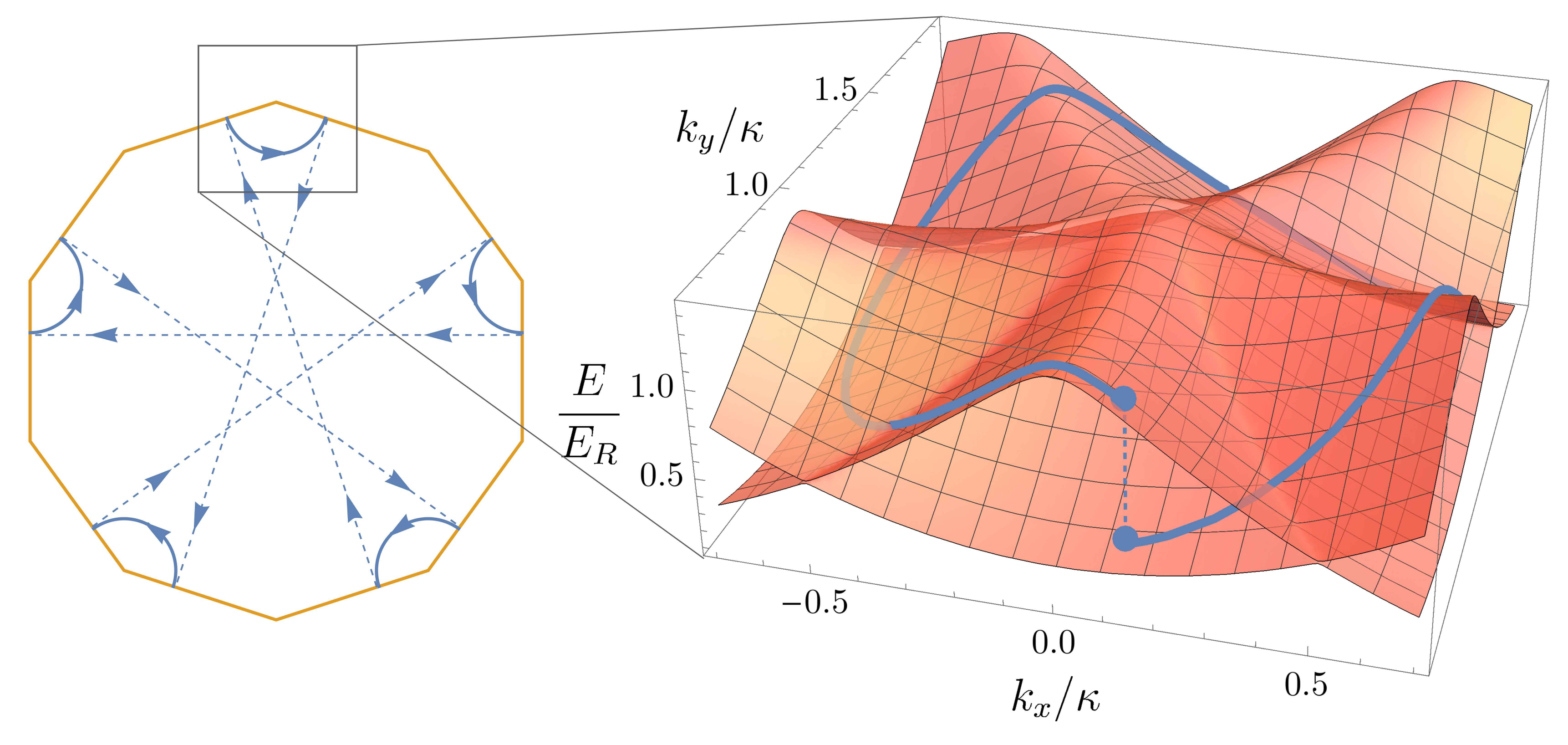}%
	\caption{A cyclic trajectory around a corner of the PBZ leads to the surprising result of a spiral holonomy \cite{cheon09new} in which after a cyclic variation of the parameter $ \kk $ the system fails to return to its initial eigenstate. This result appears in two ways: (left) the geometry of the path encircling the corner and (right) as transitions between the two lowest bands [of the Hamiltonian in Eq. (\ref{eq:Hcorner})] at a corner.}
	\label{fig:spiralholonomy}
\end{figure*}

A surprising result of semiclassical dynamics of quasicrystals in the
semiadiabatic limit is found by considering a cyclic variation of the
momentum around a corner of the PBZ. Such dynamics could be induced,
for example, by applying a force that changes in direction with time in
such a way that the net impulse imparted vanishes, such that one
expects the momentum to return to its initial value. In this case we
find that an eigenstate does not return to its original form. Instead,
the system is left in a different energy eigenstate, orthogonal to its
initial state.  (Naturally, this result will have a direct impact on
how we understand the Berry phase and Berry curvature in later
discussions.)

The origin of this phenomena can be attributed to the geometry of the PBZ. Consider following the set of Bragg scatterings, as depicted on the left of Fig.~\ref{fig:spiralholonomy}, along one cyclic path around a corner in which the momentum changes direction by $2\pi$ to encircle the corner
just once. After this single cycle, the wave packet finishes at a different corner of the PBZ.
Although the net external impulse is zero, the set of Bragg scatterings are imbalanced in such a way that there is a net momentum transfer from the quasicrystalline lattice.
It is only after performing a second $2\pi$ cycle that the particle returns to its initial location.
This unusual geometrical property manifests in the band structure local to a corner, given by Eq. (\ref{eq:Hcorner}) and as shown on the right in Fig.~\ref{fig:spiralholonomy}. This appears as a series of transitions between the two lowest bands which finishes in a different band to which it started. Such behavior is referred to as a ``spiral holonomy" \cite{cheon09new,cheon09exotic}. We emphasize that the appearance of this phenomena is a necessary consequence of working in the semiadiabatic limit for the quasicrystal.

To our knowledge, similar phenomena to what we see here---the key feature being a change in energy level after a cyclic parameter variation---have been described only in two, very different, settings for energy bands. One setting concerns the 2D surface states of a three-dimensional (3D) Weyl semimetal. Here there appears a helicoidal band structure around the projection of the Weyl point \cite{fang16topological}, that is, at the edges of the Fermi arcs of the surface metal \cite{wan11topological,xu15observation}. 
The other setting concerns  energy bands in lossy (non-Hermitian) systems. These can show  ``exceptional points'' at which the (complex) energy eigenvalue has a square root singularity between two energy levels as a function of a 2D parameter that results in the state returning to itself after two cycles \cite{heiss12physics,dembowski01experimental}. The energy level structure in both examples can be naturally thought of in terms of Riemann surfaces.


\section{Berry Phase, Berry Curvature and Chern Number\label{sec:berryphasecurvature}}

Topological and geometrical properties of the energy bands of crystalline systems are of a central interest in a large amount of fascinating recent research. Naturally some of these ideas have been extended to quasicrystalline systems \cite{kraus12topologicalstates,bandres16topological} with these works focusing on tight-binding models. Here we exploit our description based on semiclassical dynamics, to explore two fundamental quantities: the Berry phase and curvature. In the following we will focus on the properties nearby a corner of the PBZ as it is here where the Berry phase and curvature can be nonzero.

\subsection{Berry phase\label{sec:berryphase}}

The usual consideration for the Berry phase asks what geometrical phase is acquired for a cyclic parameter variation. However, as discussed in Sec.~\ref{sec:spiralholonomy}, a cyclic trajectory that encircles the corner of the PBZ returns to an orthogonal state and in this case the Berry phase cannot be defined. However, for a trajectory that encircles the corner twice, the state does return to its initial form. It is this situation which we address here.

We can find the Berry phase for a twofold trajectory by using a simple argument based on the phase acquired after a series of Bragg scatterings between the edges of the PBZ. In the local band structure picture of Fig.~\ref{fig:spiralholonomy}, as a certain state $ | \kk \ra $ adiabatically traverses an avoided crossing into a state $ | \kk' \ra $, it acquires a phase equal to that of the matrix element which opened that gap between these states, $ \la \kk' | \hat{V} | \kk \ra$. For a path that encircles the corner twice, five such adiabatic crossings are traversed---one for each scattering in Fig.~\ref{fig:spiralholonomy}---each contributing one of the five phases $ \theta_i $. Therefore the Berry phase acquired for this trajectory is given by
\begin{align}
\gamma = \sum_{i=1}^{5} \theta_i. \label{eq:berryphase}
\end{align}
A caveat to this argument is that the second-order gaps that are irrelevant far from the corner open into a first-order gap as they approach the center, as shown in Fig. \ref{fig:BCset}(c). Therefore this argument only applies to trajectories that remain sufficiently far from the corner. For the band structure shown in Fig. \ref{fig:BCset}(c) in which $ V_0/E_R = 0.3 $, a radius of approximately $ 0.2 \kappa $ would be sufficient, with this distance reducing for smaller $ V_0/E_R $.

It is important to highlight that each of the phases $ \theta_i $ in the previous argument are gauge dependent since each is equal to the phase of the matrix element $ \la \kk' | \hat{V} | \kk \ra$ which is changed by redefining the phases of the each basis element, $  | \kk \ra \to  e^{i \phi_\kk}| \kk \ra $. However, their total, $ \gamma $, is gauge invariant, as can be seen by looking at the structure of the off-diagonal couplings in (\ref{eq:Hcorner}). As shown in Fig.~\ref{fig:TotalPhase}, this set forms a closed loop in reciprocal space which ensures that any gauge transformation leaves the sum around this loop invariant.

When the Berry phase for a twofold trajectory is $ \pi $ (e.g., $ \sum_{i=1}^{5} \theta_i = \pi $), it is possible to make a connection to the physics of graphene. For graphene it is well known that the two lowest bands have a linear dispersion at Dirac points located at the corners of the Brillouin zone, each of which is associated with a $ \pi $-Berry phase. A very similar situation happens in our model---here the $ \pi $-Berry phase is also associated with a linear band touching, however, now between the second and third bands at a corner of the PBZ (this is because the lowest two bands are essentially joined by the spiral holonomy; cf. Figs. \ref{fig:spiralholonomy} and \ref{fig:BCset}). 
It is also well known that the linear dispersion (with associated $ \pi $-Berry phase) can lead to interesting phenomena such as inelastic backscattering and unusual reflection properties from a potential barrier in graphene. Since these phenomena are purely a result of this particular dispersion we expect similar phenomena to appear in our model.


\subsection{Berry curvature\label{sec:berrycurv}}

In the current section we will explore the properties of the Berry curvature of the Hamiltonian $ H^\text{corner}_\kk $ from Eq. (\ref{eq:Hcorner}) which describes mixing at a corner of the PBZ. However, first we outline some general properties of the Berry curvature based on symmetries of the system and use these ideas to derive a condition on the phases $ \theta_i $ to allow for nonzero Berry curvature. A symmetry which is present here is time-reversal symmetry, which results in $ \Omega(\kk) $ being an odd function of $ \kk $. The presence of inversion symmetry would also mean that $ \Omega(\kk) $ must be an even function of $ \kk $ and therefore both symmetries would result in zero Berry curvature. To determine whether such a point of inversion exists for the quasiperiodic potential (\ref{eq:V}), we search for  a point $ \RR $ such that
\begin{align}
V( \RR + \rr ) = V( \RR -\rr ). \label{eq:VRr}
\end{align}
It is straightforward to show that this equality is equivalent to the following set of equations:
\begin{align}
\GG_i \cdot \mathbf{R} +\theta_i = 0\ \  \text{mod}\  \pi . \label{eq:inversionequations}
\end{align}
By taking the sum of these and using the property,
\begin{align}
\sum_{i=1}^5 \GG_i = \mathbf{0},
\end{align}
one can show that the following equation must hold:
\begin{align}
\sum_{i=1}^5 \theta_i = 0 \ \ \text{mod} \ \pi. \label{eq:suminversion}
\end{align}
If this final equality fails to hold, the assumption that there exists an $ \RR $ such that $ V(\rr) $ satisfies (\ref{eq:VRr}) must be incorrect: There cannot exist a point of inversion symmetry and the Berry curvature can be nonzero.  The sum in Eq. (\ref{eq:suminversion}) is just equal to the previously found Berry phase (\ref{eq:berryphase}). Thus, the results are consistent: If the Berry phase (\ref{eq:berryphase}) is zero or $ \pi $ then the Berry curvature must be zero. The fact that the Berry phase can be equal to $ \pi $ (and therefore nonzero) while the Berry curvature must be everywhere zero is entirely analogous to the situation in graphene in which the Berry curvature is zero everywhere except at the Dirac points where it is singular.

\begin{figure}[tb]
	\centering
	\includegraphics[trim={0.0cm 0.cm 0.cm 0.cm}, clip,width=0.99\linewidth]{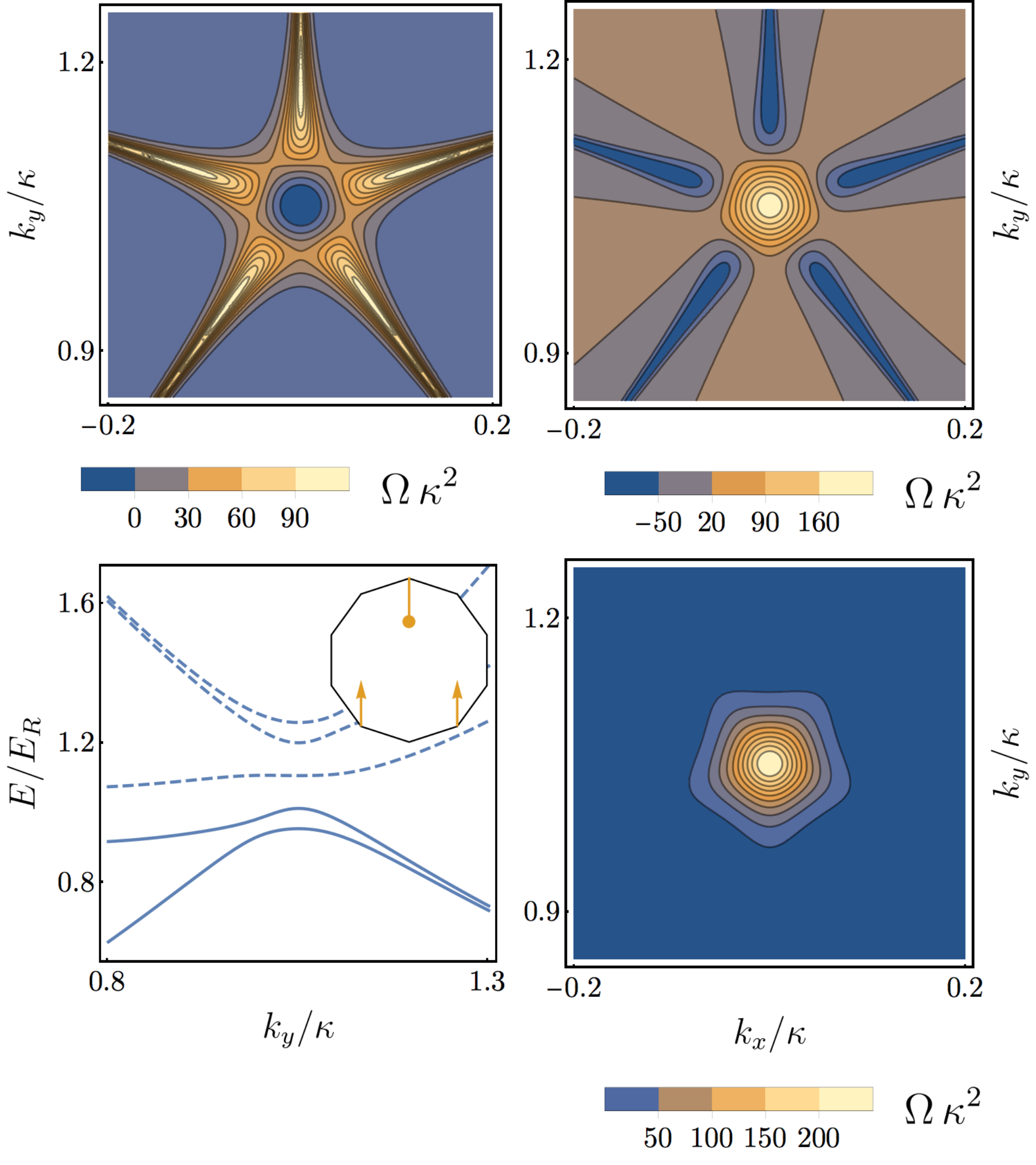}
	\caption{Berry curvature and band structure local to the corner of the PBZ for the two lowest bands of the Hamiltonian $ H^\text{corner}_\kk $ in Eq. (\ref{eq:Hcorner}), with $ V_0/E_R = 0.3 $ and $ \gamma = \pi/2 $. The separate Berry curvatures (a) $ \Omega^{(1)} $ for the first band and (b) $ \Omega^{(2)} $ for the second band, show sharp peaks along the five lines of near degeneracy shown in Fig.~\ref{fig:spiralholonomy} and in (c) the band structure past a corner. (d) For the sum $ \Omega^{(1)}+ \Omega^{(2)} $ these cancel leaving a single smooth peak which integrates to give the Berry phase associated with a twofold loop.}
	\label{fig:BCset}
\end{figure}

It is simple to find the exact form of the Berry curvatures
$ \Omega^{(n)}(\kk) $ for each of the five bands, labeled by $n$, of
(\ref{eq:Hcorner}) by using standard numerical
methods \cite{fukui05chern}. One proceeds in precisely the same way as for periodic systems (by relating the phase acquired around an infinitesimal plaquette to the curvature enclosed), the only difference for the quasicrystal is that this is carried out for an effective band structure that emerges within the semiadiabatic limit. There is, however, a subtlety here in that
calculating the Berry curvature for (\ref{eq:Hcorner}) one assumes adiabaticity with
respect to all gaps in the band structure. For the lowest band, there
are gaps of order $ V_0^2/E_R $ (cf. the discussion on the spiral
holonomy of Sec.~\ref{sec:spiralholonomy} and
Fig.~\ref{fig:spiralholonomy}), which would be tunnelled past
nonadiabatically in the semiadiabatic limit. Therefore, although
$ H^\text{corner}_\kk $ was motivated by the semiadiabatic limit, in
order to calculate the Berry curvature we must work outside of this
regime. The Berry curvature calculated here is simply that associated
with adiabatic transport for the band structure described by
$ H^\text{corner}_\kk $.

\begin{figure}[bt]
	\centering
	\includegraphics[trim={0.2cm .0cm 0.cm .0cm}, clip,width=0.99\linewidth]{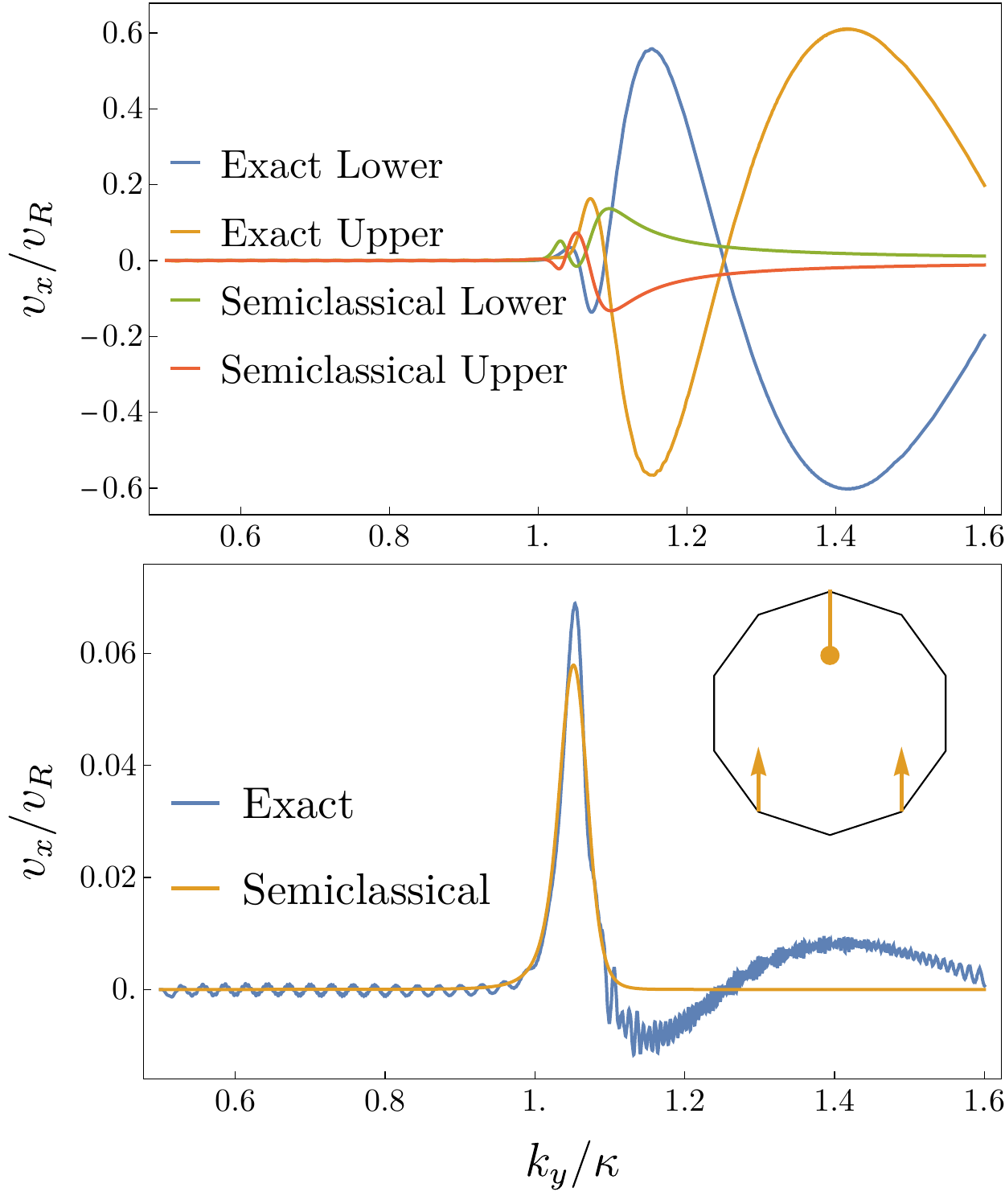}
	\caption{(Top) Due to transitions between the two lowest bands local to a corner [cf. Eq. (\ref{eq:Hcorner}) and Fig.~\ref{fig:spiralholonomy}] the separate anomalous velocities associated with the Berry curvature [Eq. (\ref{eq:SC2})] are obscured. (Bottom) Whereas their sum can be cleanly extracted (as described in the text) and matches well with the expected Berry curvature [see Fig.~\ref{fig:BCset}(c)]. The parameters used are the same as those used in Fig.~\ref{fig:SCDkx}.}
	\label{fig:SCDBC}
\end{figure}

We plot the Berry curvature of (\ref{eq:Hcorner}) for the lowest two bands as well as their sum in Fig.~\ref{fig:BCset} since generally the dynamics here will visit both bands. A striking feature of the separate Berry curvatures $ \Omega^{(1)} $ and $ \Omega^{(2)} $ are the five sharp peaks associated with the near degeneracy between the two bands. As discussed above, their relevance to the semiclassical dynamics in the semiadiabatic regime is obscured due to transitions between the bands. On the other hand, their sum $ \Omega^{(1)}+ \Omega^{(2)} $ is highly relevant within the semiadiabatic limit and can be cleanly mapped out from the semiclassical dynamics. To do so one can simply perform two evolutions, one for the particle starting in each  of the two bands and then summing the separate anomalous velocities as shown in Fig.~\ref{fig:SCDBC}. Numerically this procedure works well up to the same parameter values used in the Bloch oscillations discussion and will therefore require similar evolution times experimentally.

Aside from its appearance in the semiclassical dynamics, the Berry curvature is fundamentally related to the Berry phase via a surface integral over the region enclosed by the cyclic trajectory for which the Berry phase is defined. Making a similar statement here is subtle since for a generic trajectory one encounters transitions between the bands which means the separate adiabatic Berry curvatures are insufficient to describe the semiadiabatic Berry phase. Nevertheless for the twofold trajectory discussed in Sec.~\ref{sec:berryphase} one can associate the Berry phase here to the integral of the sum of the Berry curvatures by comparing two situations. The first in which the twofold loop is traversed semiadiabatically and a second in which two separate single loops are performed adiabatically on each band. The only difference between these two situations is found local to the near degeneracies between the two bands. In the first case no phase is acquired past these avoided crossings and in the second, while a phase is acquired for each separate band, these will cancel for the total phase from both trajectories. The result is that the semiadiabatic Berry phase $ \gamma $ acquired on a twofold trajectory [which is related to the phases $ \theta_i $ via (\ref{eq:berryphase})] is equal to the surface integral of the sum of the separate adiabatic Berry curvatures,
\begin{align}
\gamma = \iint dS (\Omega^{(1)} + \Omega^{(2)}).
\label{eq:gammaomega1omega2}
\end{align}
This result is easily confirmed numerically by integrating over the peak in the summed Berry curvatures from Fig.~\ref{fig:BCset}.

\subsection{Chern number}

Naturally one might imagine extending the surface integral of the Berry curvature in (\ref{eq:gammaomega1omega2}) over the entire PBZ, to obtain a topological invariant akin to the Chern number for the periodic case. However, one may well question whether such a topological invariant exists for the quasicrystal, since the PBZ does not have the same  topology as the BZ of conventional periodic systems which is a torus. 
Nevertheless, despite the differing topologies of the PBZ and conventional BZ, a topological invariant still exists for the PBZ. The PBZ is orientable (no subset is a M\"obius band) and closed (since all edges are identified). These two conditions of the manifold (closed and orientable) are sufficient to allow the existence of the Chern number \cite{avron85quantisation} defined by the integral of the Berry curvature over the PBZ.

Although the particular topology of the PBZ does not directly affect the Chern number, it is nevertheless interesting to ask what this topology is for the PBZ. In order to identify this, two pieces of information are needed: the orientability and the Euler characteristic $ \chi $ \cite{massey1991basic}. We already know that the PBZ is orientable (which means it is a $ g $-holed torus, where $ g $ is the genus), and the Euler characteristic is found from the number of vertices, $v$, edges, $e$, and faces, $f$, using $ \chi = v - e + f $. For the decagonal PBZ, these are $v=2$, $e=5$, $f=1$, giving $\chi=-2$, and using $ \chi =2-2g $ (for orientable surfaces) gives $ g = 2 $. We therefore identify the decagonal PBZ as a two-holed torus. Interestingly the association of a regular polygon with identified edges to a higher genus manifold also appears in the study of billiards in rational polygons \cite{zorich2006flat}. There the straight line billiard trajectories are interpreted as curved trajectories on this manifold. Surprisingly this situation is closely related to the straight line $ k $-space trajectories in our model for constant external force (cf. Sec \ref{sec:blochiscillations}).


\section{Generalizations\label{sec:Generalisations}}

\subsection{Semiclassical dynamics in solid state quasicrystals}

The semiclassical approach we have presented in Sec.~\ref{sec:SCD} is very general. The only assumption it relies on is that the hierarchy of gaps can be clearly separated in terms of their sizes. For this condition to be satisfied two criteria must be met: the first is that the Fourier components of the potential must fall off sufficiently quickly (in our case only ten were nonzero). The second is that these components must also be sufficiently weak so that higher order effective couplings can be neglected (here this meant working in the shallow-lattice limit). Both conditions can be satisfied in an optical lattice setting, since the potentials are often formed by a small number of standing waves and the lattice depth is freely tunable. 

Surprisingly these conditions could also be satisfied for a solid-state quasicrystal, as a number of ARPES studies on various icosahedral and decagonal solid-state quasicrystals have demonstrated that these have a free-electron-like dispersion \cite{rotenberg2000quasicrystalline,theis03electronic,rogalev2015fermi}. 
Of course disorder plays a key role in these materials, likely obscuring the semiclassical dynamics. However, there are situations---like in quantum oscillations---where semiclassical dynamics remain highly relevant. Indeed, related  ideas to those presented here were already used in Ref. [\onlinecite{zhang15disruption}] to explain quantum oscillations in incommensurate charge density waves. The nature of the quantum oscillations in our model presents an interesting open question, the answer to which could be of relevance to the properties of icosahedral and decagonal solid-state quasicrystals which share the same rotational symmetry.

\subsection{Higher rotational symmetries}

Many of the results presented here can be simply extended to systems with arbitrary rotational symmetries. These include the spiral holonomy, the possibility of nontrivial Berry phases and curvature, and the identification of a Chern number. Essentially these only depend on the overall geometry of the PBZ, so that as long as a PBZ can be well defined one can ask such questions. We discuss generalized PBZ's which are regular $ 2n $-sided polygons, with integer $ n \ge 4 $. The results will naturally split into two cases for odd or even $ n $. With the model studied throughout this paper given by $ n=5 $ and therefore an odd case. 

For the spiral holonomy, the same geometrical picture used in Sec.~\ref{sec:spiralholonomy} and shown in Fig.~\ref{fig:spiralholonomy} to find the number of cycles around a corner before returning can be applied here. For odd $ n $, the trajectory visits only $ n $ of the total $ 2n $ corners before returning and therefore completes $ (n-1)/2 $ cycles (e.g., in our case $ n=5 $ and $ 2 $ cycles were required). Whereas for even $ n $, the trajectory visits all $ 2n $ corners, resulting in a total of $ n-1 $ cycles before returning to the initial state. For example, if $ n=4 $ (e.g. an octagonal PBZ as in Fig. 9), the state will require three cycles around a corner before returning and will therefore visit three bands local to the corner. Three cycles also implies a chirality, since going clockwise or anticlockwise produces different results. 

An interesting difference between odd and even $ n $ appears by asking whether one can find nonzero Berry curvature. The odd case is essentially the same as the fivefold case in this respect. Half the corners are coupled in such a way that the off-diagonal terms again form a closed loop allowing for nonzero Berry curvature. However in the even case, all $ 2n $ corners couple (cf. Fig. 9), forcing the Berry curvature to be the same at all corners. This is related to how the state visits all $ 2n $ corners in the spiral holonomy. If time-reversal symmetry is present, $ \Omega(\kk) $ must be an odd function of $ \kk $, and the only possible Berry curvature at a corner is zero. Therefore for even $ n $ it is not possible to have nontrivial Berry phases or curvature while time-reversal symmetry is preserved.

Finally, the Chern number classification can be easily extended, since for all $ n $ the PBZ is both orientable and closed, and therefore the Chern theorem applies. The genus can then be found by calculating the Euler characteristic. For odd $ n $, the PBZ is found to have genus $ (n-1)/2 $, while for even $ n $ it has genus $ n/2 $. With the difference between odd and even cases again arising from how the corners are coupled---for odd $ n $ there are two vertices while for even $ n $ there is only one vertex. Therefore for all $ n $, integrating the Berry curvature over the whole PBZ provides a topological invariant---the Chern number.

\begin{figure}[tb]
	\centering
	\includegraphics[trim={0.0cm 0.cm 0cm 0.cm}, clip,width=0.63\linewidth]{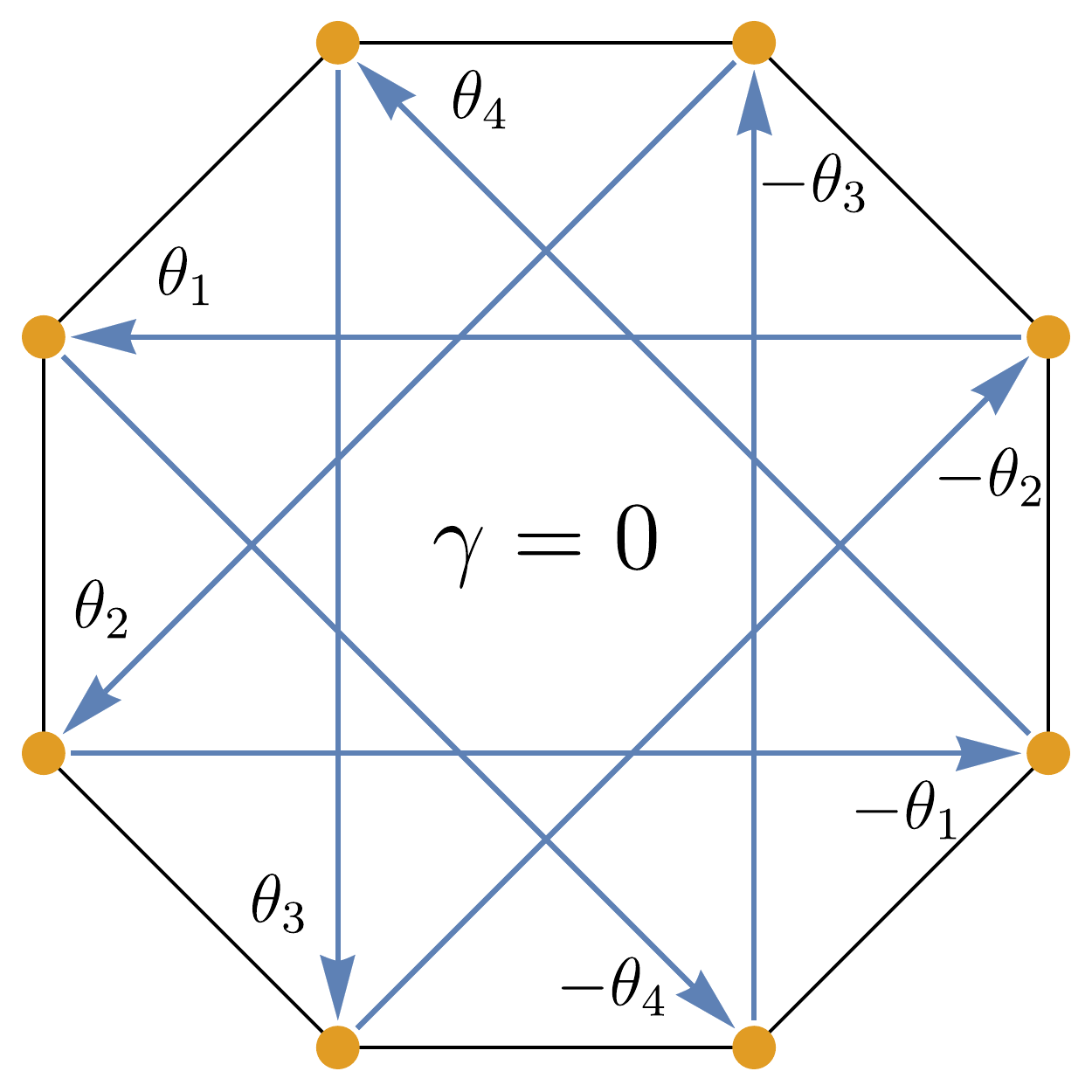}
	\caption{Same as Fig.~\ref{fig:TotalPhase} but for a PBZ with eight sides. Here the couplings again form a closed loop meaning the total phase is gauge invariant, however, the sum is now zero and therefore the Berry phase and curvature are also zero.}
	\label{fig:TotalPhase8fold}
\end{figure}

	
\section{Conclusion}

We have demonstrated that for a two-dimensional shallow-lattice optical quasicrystal, it is possible to identify a regime in which the dynamics is accurately described by the semiclassical equations of motion. By comparing the prediction of Bloch oscillations against an exact numerical solution we determined the maximum potential depth allowed in order for the semiclassical description to apply and related this to experimental parameters. 

A surprising result was the appearance of a spiral holonomy around a corner of the PBZ---a phenomena which has been described in only a few, very different, settings for energy bands. We also demonstrated that it is possible to have nontrivial Berry phase and curvature at a corner---with both having an unconventional structure due to the spiral holonomy. 
A method of extracting the Berry curvature from the semiclassical dynamics was provided and its overall properties were related to time-reversal and inversion symmetries. 
By identifying the PBZ as topologically equivalent to a higher genus surface, we showed that the Chern number classification for periodic systems can be extended to the PBZ of a quasicrystal, thereby determining a topological index for the system.

We highlight that the semiclassical approach can be applied to a generic quasicrystal
and can be applicable in solid-state quasicrystals with a
nearly-free-electron dispersion which have been observed
experimentally. We have also extended the findings of the spiral holonomy,
Berry curvature, and Chern number to systems with arbitrary rotational symmetries by
relating these to the properties of the PBZ. We show that Berry curvature effects appear for certain ``odd'' arrangements but
disappear for ``even'' arrangements.

\acknowledgements{We gratefully acknowledge helpful discussions with Ulrich Schneider and Claudio Castelnovo, along with helpful suggestions from two anonymous referees and support from  the EPSRC via Grants No. EP/K030094/1 and No. EP/P009565/1
and the Simons Foundation.
Statement of compliance with EPSRC policy framework on research data: All data are directly available within the publication.}

\bibliography{../../library}

\end{document}